\documentclass[11pt,a4paper]{article}
\usepackage{amsmath,amssymb}
\usepackage{pstricks}
\usepackage{braket}
\usepackage{pst-plot}
\usepackage{amscd,float,array,bm}
\usepackage{graphicx}
\usepackage{epsfig}

\usepackage{braket}

\newcommand{\Ree}{{\rm Re}}
\newcommand{\Ha}{{\rm Ha}}
\newcommand{\Rm}{{\rm Rm}}
\newcommand{\Pm}{{\rm Pm}}

\newcommand{\er}{{\bf\hat e}_r}
\newcommand{\et}{{\bf\hat e}_\theta}
\newcommand{\ep}{{\bf\hat e}_\varphi}
\newcommand{\ez}{{\bf\hat e}_z}
\newcommand{\ve}{{\mathbf{v}}}
\newcommand{\bv}{{\bf B}}

\newcommand{\lp}{\ensuremath{\left(}}
\newcommand{\rp}{\ensuremath{\right)}}

\begin{document}

\title{Continuation and stability of rotating waves in the magnetized
  spherical Couette system: Secondary transitions and multistability}

\author{F. Garcia$^{1,2}$ and F. Stefani$^1$\\
  {\small{1 Helmholtz-Zentrum Dresden-Rossendorf}}\\{\small{, P.O. Box 510119, D-01314 Dresden, Germany.}}\\
  {\small{2 Anton Pannekoek Institute for Astronomy, University of Amsterdam,}}\\ {\small{Postbus 94249, 1090 GE Amsterdam, The Netherlands.}}\\}







\maketitle

\begin{abstract}
Rotating waves (RW) bifurcating from the axisymmetric basic magnetized
spherical Couette (MSC) flow are computed by means of Newton-Krylov
continuation techniques for periodic orbits. In addition, their
stability is analysed in the framework of Floquet theory. The inner
sphere rotates whilst the outer is kept at rest and the fluid is
subjected to an axial magnetic field. For a moderate Reynolds number
$\Ree=10^3$ (measuring inner rotation) the effect of increasing the
magnetic field strength (measured by the Hartmann number $\Ha$) is
addressed in the range $\Ha\in(0,80)$ corresponding to the working
conditions of the HEDGEHOG experiment at Helmholtz-Zentrum
Dresden-Rossendorf. The study reveals several regions of
multistability of waves with azimuthal wave number $m=2,3,4$, and
several transitions to quasiperiodic flows, i.e modulated rotating
waves (MRW). These nonlinear flows can be classified as the three
different instabilities of the radial jet, the return flow and the
shear-layer, as found in previous studies. These two flows are
  continuously linked, and part of the same branch, as the magnetic
forcing is increased. Midway between the two instabilities, at a
certain critical $\Ha$, the nonaxisymmetric component of the flow is
maximum.
\end{abstract}


\section{Introduction}

The origin of the magnetic fields of planets, stars and galaxies
constitutes one of the most challenging problems of modern
physics. Larmor was the first to suggest that the magnetic fields of
the Sun, the Earth and cosmic bodies are supported by electrically
conducting fluid motions in their interiors \cite{Lar19}. Many years
later, the first successful dynamo experiments with liquid
sodium~\cite{GLPGS02} supported this idea. The dynamo problem involves
a large range of scales, and hence provides tremendous experimental,
analytical and numerical challenges in the parameter regime relevant
to geophysical and astrophysical applications.  We refer to the review
article~\cite{BrSu05} and a book~\cite{DoSo07} for detailed references
and the history of the field.

One of the paradigms of magnetohydrodynamic flows in spherical bodies
such as planets and stars is magnetized spherical Couette (MSC)
flow. An electrically conducting liquid is confined between two
differentially rotating spheres and is subjected to a magnetic field.
Despite its simplicity, this model gives rise to a rich variety of
instabilities, and it is also important from an astrophysical point of
view. For instance, simulations of spherical Couette (SC) flow were
used to compute the gravitational wave signal generated by global
nonaxisymmetric shear flows in a neutron star~\cite{PMGO06}. In
addition, instabilities observed in a liquid sodium flow between
differentially rotating spheres in the presence of a magnetic field
were attributed in~\cite{SMTHDHAL04} to the magnetorotational
instability (MRI), but see~\cite{Hol09,GJG11,KKSS17} for alternative
interpretations. Since the pioneering work of~\cite{BaHa91}, MRI has
been considered the best candidate to explain the mechanism of
transporting angular momentum in accretion disks around black holes
and stars, and also in protoplanetary disks~\cite{JiBa13}, allowing
matter to fall into the center.  Various types of the MRI have also
been studied experimentally at Helmholtz-Zentrum Dresden-Rossendorf
(HZDR) \cite{SGGRSSH06,SGGHPRS09,SGGGSGRSH14}.

Although the MSC represents one of the simplest paradigms of
astrophysical magnetohydro-dynamics, it possesses several
peculiarities that make the problem difficult. From the analytical
point of view, the nonlinear nature of the Navier-Stokes equations and
the spherical geometry of the boundaries lead to mathematical
complications for developing analytical theories that are indeed hard
to treat. For this reason, the development and improvement of
appropriate numerical techniques is of key importance for a deep
understanding of nonlinear flows, even in the weakly supercritical
regime. In addition, thin Ekman or Ekman-Hartmann boundary layers,
depending on the strength of the magnetic field, appear when the
no-slip condition, used to model planetary dynamos and for comparison
with laboratory experiments, is imposed at one boundary. Even in the
absence of a magnetic field,~\cite{Ste66} showed analytically the
existence of a thin shear layer (Stewartson layer) at the tangent
cylinder (containing the inner sphere and parallel to the rotation
axis) which separates regions of different flow behavior. These thin
shear layers make the numerical treatment extremely challenging
because of the higher spatial resolution which is computationally
most demanding.

In the absence of magnetic fields, the solutions of the SC problem
including the basic flow, the first instabilities, and even turbulent
states have been widely studied experimentally~\cite{NaTs95,ZTL11},
analytically~\cite{Ste66,MuJo75} and numerically using direct
numerical simulations (DNS)~\cite{Zik96,HJE06} or continuation
methods~\cite{Sch86,MaTu95}. Many fewer numerical studies exist
in which the magnetic field is taken into account. Most studies deal
with the linear stability analysis of the basic flow, or are build on
the basis of very few nonlinear solutions, and mainly rely on
considering different types of boundary conditions for the magnetic
field~\cite{HoSk01,GJG11} (insulating or conducting inner sphere
allowing magnetic lines to pass), different topologies of the applied
magnetic field (dipolar~\cite{GJG11,FSNS13},
axial~\cite{SMTHDHAL04,Hol09,Kap14}, or a combination of
both~\cite{WeHo10}). Sophisticated tools of hydrodynamic stability
theory, such as continuation techniques, are in many aspects superior
to simple DNS. For instance, time integration methods are unable to
obtain unstable oscillatory solutions when all the symmetries of the
flow are broken. These solutions might be relevant in organizing the
global dynamics~\cite{KUL12}. Bifurcation and continuation methods
have been successfully applied during the last years to a great
variety of problems in the fluid dynamics
context~\cite{KUL12,DWCDDEGHLSPSSTT14}. Computations based on
continuation of periodic orbits of nontrivial time
dependence~\cite{Vis07,GNS16} and even tori~\cite{SNS10} or other
invariant objects~\cite{VKA11} have provided useful information to
clarify the dynamics.

The solution (basic state) of the SC equations is unconditionally
stable up to a certain critical value of the forcing parameter (the
Reynolds number, $\Ree$, measuring the strength of differential
rotation). Beyond this threshold an instability develops and a branch
of stable or unstable solutions bifurcates and extends into a certain
region of the parameter space. The appearance of multiple states in
experimental flows strongly depends on the initial
state~\cite{NaTs95}, and in DNS the type of perturbation applied
determines the type of instability or the mode that will be selected
among the bifurcated solutions~\cite{HJE06}. To get a complete picture
of the skeleton of the phase space, and thus to provide a better
characterization of the instabilities and the physically realizable
flows, a continuation method is
necessary~\cite{Kel77,MaTu95,SGN13,GNS16} (see~\cite{SaNe16} for a
nice tutorial).

Due to the spherical geometry and rotation the SC system has
$\mathcal{SO}(2)$ symmetry. Therefore, the instability of the basic
flow usually gives rise to waves traveling in the azimuthal direction,
i.e. rotating waves (RW), which break the axisymmetry of the basic
state~\cite{GoSt03}. A secondary Hopf bifurcation results in an
amplitude or shape modulation of the flow pattern, i.e. the occurrence
of modulated rotating waves (MRW), which may have different types of
spatio-temporal symmetries~\cite{Ran82,GLM00}. In the somewhat
different case of the Taylor-Couette system, several types of RW and
MRW have been identified and characterized depending on their
symmetry~\cite{ALS86}. In spherical geometry Schaeffer et
al.~\cite{ScCa05} have shown that the destabilization of the
Stewartson layer, which is characteristic for the basic state, gives
rise to a Rossby wave of fixed azimuthal wave number, travelling in
the azimuthal direction due to the curvature of the boundaries. In
addition, by means of fully three-dimensional
simulations,~\cite{HJE06} reported further transitions in the
supercritical regime in which the original azimuthal symmetry is
replaced by a so-called shift-and-reflect
symmetry~\cite{Kuz98,GoSt03}. In case of the MSC problem the existence
of RW and MRW has been confirmed by experimental
studies~\cite{SABCGJN08}, and by DNS~\cite{HoSk01,Hol09,GJG11}.

Addressing the influence of the applied magnetic field, ~\cite{HoSk01}
have shown that the axisymmetric basic state of the axially MSC
problem is equatorially symmetric and remains stable for all Hartmann
numbers $\Ha$ (measuring the strength of the applied magnetic field)
if the Reynolds number is sufficiently small. It can be described as a
strong azimuthal flow associated with a meridional recirculation. As
the Reynolds number is increased, the basic state becomes unstable to
non-axisymmetric perturbations. At low Hartmann number these
perturbations are equatorially antisymmetric giving rise to an
instability which is essentially hydrodynamic and related to a
Kelvin-Helmholtz instability (KHI) of the radial jet at the equatorial
plane. MSC is also $\mathcal{Z}_2$, i.e, invariant by reflections with
respect to the equatorial plane, but the nonlinear saturation of
  the radial jet instability is equatorially asymmetric. At
sufficiently large Hartmann numbers the perturbations become
equatorially symmetric. For small rotation rates this instability is
related to a shear layer at the tangent cylinder~\cite{HoSk01,Hol09}
while at higher rotations the instability is located at the base of
the meridional return flow~\cite{Hol09}. In this case, increasing
further the Reynolds number stabilizes the basic
flow~\cite{Hol09,TEO11}.  When holding the Reynolds number fixed,
which is the approach of the present study and of the preliminary
experiments performed in~\cite{KKSS17}, the two types of instability
are separated by a stable regime which occurs for intermediate
Hartmann numbers.

Our analysis partially fills the gap between the very high Reynolds
number turbulent regime reached in some of the
experiments~\cite{SMTHDHAL04} and some numerical studies~\cite{FSNS13}
and the low Reynolds number laminar regime in which the linear
stability of the basic state has been deeply analyzed but the
nonlinear saturation of the instabilities has only been studied using
a few nonlinear simulations~\cite{HoSk01,Hol09,TEO11}. The use
  of continuation techniques allows us to obtain precise bifurcation
  diagrams and to determine the stability regions of RW with
azimuthal wave number $m=2,3,4$ in the range of $\Ha\in(0,80)$. The
paper is organized as follows: In \S\ \ref{sec:mod} we introduce the
formulation of the problem, and the numerical method used to integrate
the model equations. Next, the continuation method and the basic
ingredients for the stability analysis are briefly described in
\S\ \ref{sec:co_st_wv}. In \S\ \ref{sec:res} the bifurcation diagrams
as a function of $\Ha$, the stability of RW and the patterns of
convection are analyzed.  Finally, in \S\ \ref{sec:sum} the paper
closes with a brief summary on the results obtained.

\section{The Model}
\label{sec:mod}

Let us consider a spherical shell of inner and outer radii $r_i$ and
$r_o$. The outer sphere is at rest while the inner rotates at a
constant angular velocity $\Omega$ around the $\ez$ axis. The shell is
filled with a homogeneous and conducting fluid of constant density
$\rho$, dynamic viscosity $\mu$, magnetic diffusivity $\lambda$ and
electrical conductivity $\sigma=1/(\lambda\mu_o)$, where $\mu_0$ is
the free-space value for the magnetic permeability.

We are interested in a comparison with laboratory
experiments~\cite{KKSS17} which subject the flow to a uniform axial
magnetic field $\bv_0=B_0 \cos(\theta)\er-B_0 \sin(\theta)\et$,
$\theta$ being the colatitude and $B_0$ the magnetic field
strength. With the use of the eutectic alloy GaInSn as the working
fluid of the HEDGEHOG experiment, the inductionless approximation can
be adopted.  This approximation is valid in the limit of low
magnetic Reynolds number $\Rm=\Pm\Ree \ll 1$, which applies in the
case of the HEDGEHOG experiment because of its very low
magnetic Prandtl number fluid (GaInSn) with $\Pm\sim O(10^{-6})$ and
the moderate Reynolds numbers (inner sphere rotation rate)
considered $\Ree\sim 10^3$.

By scaling length, time, velocity and magnetic field by $d=r_o-r_i$,
$d^2/\nu$, $r_i\Omega$ and $B_0$, respectively, decomposing the
magnetic field as $\bv=\ez+\Rm{\bf b}$ and neglecting terms $O(\Rm)$,
the Navier-Stokes and induction equations become
\begin{align}
  \partial_t\ve+\Ree\lp\ve\cdot\nabla\rp\ve &=
-\nabla p+\nabla^2\ve+\Ha^2(\nabla\times {\bf b})\times\ez, \label{eq:mom_less}   \\
 0& = \nabla\times(\ve\times\ez)+\nabla^2{\bf b}, \label{eq:ind_less}\\
\nabla\cdot\ve=0, &\quad \nabla\cdot{\bf b}=0.\label{eq:div}
\end{align}
In this inductionless approximation the system is governed by only
three non-dimensional numbers, namely the Reynolds number, the
Hartmann number and the aspect ratio:
\begin{equation*}
  \Ree=\frac{\Omega r_i d}{\nu}, \quad
  \Ha=\frac{B_0d}{\sqrt{\mu_0\rho\nu\lambda}}=B_0d\sqrt{\frac{\sigma}{\rho\nu}},\quad
  \eta=\frac{r_i}{r_o}.
\end{equation*}
No-slip ($v_r=v_\theta=v_\varphi=0$) at $r=r_o$ and constant rotation
($v_r=v_\theta=0,~v_\varphi= \sin{\theta}~ \ep$) at $r=r_i$ are the
boundary conditions imposed on the velocity field. For the
magnetic field, insulating exterior regions are considered in
accordance with the experimental setting, see~\cite{HoSk01} for more
details.

The equations are discretized and integrated with the same method as
described in~\cite{GNGS10} and references therein.  The velocity and
magnetic fields are expressed in terms of toroidal and poloidal
potentials and are expanded in spherical harmonics in the angular
coordinates, and in the radial direction a collocation method on a
Gauss--Lobatto mesh is used.  The code is parallelized in the spectral
and in the physical space by using OpenMP directives. We use optimized
libraries (FFTW3~\cite{FrJo05}) for the FFTs in $\varphi$ and
matrix-matrix products (dgemm GOTO~\cite{GoGe08}) for the Legendre
transforms in $\theta$ when computing the nonlinear terms.

For the time integration, high order implicit-explicit backward
differentiation formulas (IMEX--BDF)~\cite{GNGS10} are used. In the
IMEX method we treat the nonlinear terms explicitly in order to avoid
solving nonlinear equations at each time step. The Lorenz term is
also treated explicitly, which may reduce the time step in
comparison with an implicit treatment. However, this is not a serious
issue when moderate $\Ha$ are considered, as is the case of the
present study.  The use of \textit{matrix-free} Krylov methods
(GMRES~\cite{SaSc86} in our case) for the linear systems facilitates
the implementation of a suitable order and time stepsize control for
the time integration (see~\cite{GNGS10} for details on the
implementation).

\section{Computation and stability of RW}
\label{sec:co_st_wv}

The system of Eqs.~(\ref{eq:mom_less}-\ref{eq:div}) is
$\mathcal{SO}(2)\times\mathcal{Z}_2$-equivariant, $\mathcal{SO}(2)$
generated by azimuthal rotations, and $\mathcal{Z}_2$ by reflections
with respect to the equatorial plane.  According to bifurcation
theory~\cite{EZK92,GLM00}, the first bifurcation, which breaks the
axisymmetry of the basic state, is a Hopf bifurcation giving rise to a
rotating wave (RW). The linear stability analysis of the basic
state~\cite{TEO11} provides the critical values $\Ha_c$ and the
drifting frequencies $\omega_c$ of these nonaxisymmetric instabilities
as a function of $\Ree$ and $\eta$.

Rotating waves, $u(r,\theta,\varphi-\omega t)=\tilde
u(r,\theta,\tilde\varphi)$, with $\tilde\varphi=\varphi-\omega t$, can
be obtained efficiently by Newton-Krylov continuation methods as
steady solutions of the equations written in a reference frame which
is rotating with the wave, see for instance~\cite{SGN13} for thermal
convection in spherical geometries or~\cite{TLW19} for pipe
flow. In~\cite{SGN13} RW were obtained as a zeros of a nonlinear
system, requiring the use of preconditioning techniques, to accelerate
the convergence of the linear solver associated to the Newton's
method. In contrast, the Stokes preconditioning method used
in~\cite{TLW19} only relies on time integration of the large nonlinear
system over a single time step and thus is easier to implement if a
time stepping code is already available. However, as mentioned
in~\cite{GNS16}, RW can also be found as periodic orbits which is the
approach followed in the present study. We note this method is
  not as efficient in large scale
  $\mathcal{SO}(2)\times\mathcal{Z}_2$-equivariant systems as those
  proposed in~\cite{SGN13,TLW19}.

Some background necessary to follow easily this section is now
provided. The discretization of Eqs.~(\ref{eq:mom_less}-\ref{eq:div})
leads to a system of ordinary differential equations (ODE) of
dimension $n=(2L_{\text{max}}^2+4L_{\text{max}})(N_r -1)$,
$L_{\text{max}}$ and $N_r$ being the spherical harmonic
truncation parameter and the number of radial collocation points,
respectively. The ODE system takes the form
\begin{align}
L_0\partial_tu=L u+B(u,u),
\label{eq:ode_eq}
\end{align}
where $u$ is the vector containing the spherical harmonic coefficients
of the velocity potentials at the inner radial collocation points, and
$L_0$ and $L$ are linear operators which include the boundary
conditions, $L_0$ being invertible.  The operator $L$ includes all the
linear terms and depends on the Hartmann number $\Ha$, which will be
the control parameter of this study. The other parameters are
kept fixed to $\eta=0.5$ and $\Ree=10^3$. Therefore, $p=\Ha$ and
$L=L(p)$. The bilinear operator $B$ contains the non-linear
(quadratic) terms.

\subsection{Continuation of RW}
\label{sec:co_wv}

To study the dependence of RW, rotating at a frequency $\omega$ and
with $m$-fold azimuthal wave number, on the parameter $p=\Ha$,
pseudo-arclength continuation methods for periodic orbits are
used~\cite{SNGS04b,SaNe16}. These methods obtain the curve of
periodic solutions $x(s)=(u(s),\tau(s),p(s))\in\mathbb{R}^{n+2}$, $s$
being the arclength parameter and $\tau=2\pi/(m\omega)$ the rotation
period, by adding the pseudo-arclength condition
\begin{equation*}
m(u,\tau,p)\equiv\langle w,x-x^0 \rangle=0,
\end{equation*}
where $x^0=(u^0,\tau^0,p^0)$ and $w=(w_u,w_\tau,w_p)$ are the
predicted point and the tangent to the curve of solutions,
respectively, obtained by extrapolation of the previous points along
the curve.

The system which determines a single solution,
$x=(u,\tau,p)$ is
\begin{equation}
H(u,\tau,p)= \left(
\begin{array}{c}
u-\phi(\tau,u,p)\\
g(u)\\
m(u,\tau,p)\\
\end{array}
\right)
=0,
\label{eq:H_eq}
\end{equation}
where $\phi(\tau,u,p)$ is a solution of Eq.~(\ref{eq:ode_eq}) at
time $\tau=2\pi/(m\omega)$ and initial condition $u$ for fixed
$p$. The condition $g(u)=0$ is selected to fix the undetermined phase
of the RW.  We use $g(u)=\langle u,\partial_{\varphi}u_c \rangle$
where $u_c$ is a reference solution (a previously computed solution,
or a eigenfunction provided by the linear stability analysis, for
instance). It is a necessary condition for $\|u-u_c\|^2_2$ to be
minimal with respect to the phase (see~\cite{SGN13}). For the
computation of the inner products $\langle\cdot,\cdot\rangle$ between
two functions expanded in spherical harmonics we use the definitions
of~\cite{SGN13}.

To solve the large non-linear system defined by Eq.~(\ref{eq:H_eq}) we
use Newton-Krylov methods. These matrix-free methods do not
require the explicit computation of the Jacobian
$D_{(u,\tau,p)}H(u,\tau,p)$, but only its action on a given
vector. For the linear systems we use GMRES~\cite{SaSc86}. Due to the
particular form of the spectrum of $D_{(u,\tau,p)}H(u,\tau,p)$ for
dissipative systems, GMRES does not need preconditioning
(see~\cite{SNGS04b} for details).

The action of the Jacobian $D_{(u,\tau,p)}H(u,\tau,p)$
on $\delta x=(\delta u,\delta \tau, \delta p)\in\mathbb{R}^{n+2}$ is
\begin{equation*}
\left(
\begin{array}{c}
\delta u-v(\tau)-\dot{z}(\tau)\delta \tau\\
D_ug(u)\delta u \\
D_xm(x)\delta x \\
\end{array}
\right)\in\mathbb{R}^{n+2}.
\end{equation*}
Here $z(\tau),v(\tau)\in\mathbb{R}^{n}$ are the solutions, at time
$t=\tau$, of the system
\begin{align*}
&\partial_tz=L_0^{-1}(L(p)z+B(z,z)) ,\\
  &\partial_tv=L_0^{-1}(L(p)v+B(z,v)+B(v,z)) +2 p \delta p L_0^{-1}L^{(2)}z,
\end{align*}
with initial conditions $z(0)=u$ and $v(0)=\delta u$, with fixed
$p$. The dependence of $L$ on $p$ has been assumed to be of the form
$L(p)=L^{(1)}+p^2 L^{(2)}$. Each GMRES iteration will require one
evaluation of the Jacobian, therefore most of the computational cost
is consumed in the integration over one tentative rotation period
$\tau$ of a large ODE system of dimension $2n$. An efficient
time-stepping code is hence a key ingredient for a successful
application of the method.

\subsection{Stability of RW}
\label{sec:st_wv}

Suppose a RW $(u,\tau,p)\in\mathbb{R}^{n+2}$ has been found (we recall
$\tau=2\pi/(m\omega)$). To study the stability of this periodic
solution, Floquet theory is applied. Handling the full Jacobian matrix
$D_u \phi(\tau,u,p)$, where $\phi(\tau,u,p)$ is the solution of
Eq.~(\ref{eq:ode_eq}) at time $t=\tau$ with initial condition $u$ and
for fixed $p$, would require a prohibitive amount of memory due the
high resolution employed in the present study. Fortunately, it is
enough to compute the dominant eigenvalues and eigenvectors of the map
$\delta u\longrightarrow D_u \phi(\tau,u,p)\delta u= v(\tau)$, with
$v(\tau)$ being the solution of the first variational equation,
obtained by integrating the system
\begin{align*}
&\partial_t z=L_0^{-1}(L(p)z+B(z,z)),\\
&\partial_t v=L_0^{-1}(L(p)v+B(z,v)+B(v,z)), 
\end{align*}
of dimension $2n$, with initial conditions $z(0)=u$ and $v(0)=\delta
u$, over a rotation period $\tau$, with fixed $p$.

The eigenvalues of the map with larger modulus, which correspond to
the dominant Floquet multipliers, are computed by using the ARPACK
package. RW with dominant Floquet multipliers with modulus larger
(smaller) than $+1$ are unstable (stable). Note that in this problem,
for any value of $p$, there is a marginal ($+1$) Floquet multiplier
due to the invariance under azimuthal rotations, with associated
eigenfunction $v_1=\partial_t u$. To avoid unnecessary computations it
can be deflated by computing the eigenvalues of the map $\delta u
\longrightarrow v(\tau)-\braket{v(\tau),v_1}v_1$.  This method to
determine the stability of the solutions is very robust but
computationally expensive because it requires the time integration of
an ODE system of dimension $2n$ over one rotation period. Because the
periodic orbit is a RW there is a more efficient alternative to this
procedure~\cite{SGN13,Tuc15} which consist of studying the stability
as a fixed point of a vector field. Matrix transformations could be
used (real shift invert, Cayley, etc.) to extract the right-most
eigenvalues of the associated eigenvalue problem. Preconditioning
techniques, readily available from the spatial discretization, are
used to accelerate the convergence of the iterative solver.

\subsection{Validation of the results}
\label{sec:val_res}

\begin{table}[t!] 
  \begin{center}
\scalebox{0.86}{\begin{tabular}{lcccccccc}
\vspace{0.1cm}            
$N_r$  & $L_{\text{max}}$  & $m$   & Eq. sym & $\Ha$       & $\omega$    & $|\lambda|$   & $\text{Arg} \lambda$ & $n$ \\
\hline\\
$40$ & $84$  & $3$   & $0$    & $3.1316477$ & $139.3432$ & $1.0463643$ & $2.4932338$  & $563472$   \\
$60$ & $126$ & $3$   & $0$    & $3.1316477$ & $138.9097$ & $1.0502121$ & $2.5010735$  & $1903104$  \\
$40$ & $84$  & $4$   & $1$    & $27.327395$ & $71.12972$ & $0.9079801$ & $2.2812436$  & $563472$   \\
$60$ & $128$ & $4$   & $1$    & $27.327395$ & $71.12765$ & $0.9080306$ & $2.2811844$  & $1963520$  \\
\hline
  \end{tabular}}
  \caption{Spatial discretization study. Frequency $\omega$ and
    dominant complex Floquet multiplier,
      $\lambda=|\lambda|e^{i\text{Arg} \lambda}$, as function of the
    number of radial collocation points $N_r$ and the spherical
    harmonics truncation parameter $L_{\text{max}}$. The total number
    of degrees of freedom of the system
    $n=(2L_{\text{max}}^2+4L_{\text{max}})(N_r-1)$ is also shown.}
  \label{table:mesh_study}
  \end{center}
\end{table}

Throughout the study several numerical tests, that we summarise in the
following, have been performed to validate the results. For the
continuation of the waves as well as the computation of its stability
the fixed step time integration is checked with an VSVO implicit
method and very low tolerances (see~\cite{GNGS10} for details on the
time integration methods). Moreover, changing tolerances, Krylov
dimension or number of desired eigenvalues when dealing with Arnoldi
methods (ARPACK) helps to identify changes of stability. An initial
random seed is integrated over several rotation periods before calling
the eigenvalue solver. This improves the convergence to the dominant
eigenfunctions by filtering spurious modes~\cite{SGN16}. In addition,
the spatial resolution is changed from time to time to look for
discretization errors, see table~\ref{table:mesh_study}. On this
  table, the values of the drifting frequency and the modulus and
argument of the dominant Floquet multiplier are shown for two
different solutions belonging to the equatorially asymmetric and
symmetric regions, respectively. Errors below $1\%$ are obtained when
increasing the resolution from $N_r=40$, $L_{\text{max}}=84$ to
$N_r=60$, $L_{\text{max}}=128$. Notice that for the higher
resolution the number of degrees of freedom rises to nearly $2\times
10^6$.  Although DNS studies of this problem have been carried out for
resolution as high as $N_r=60$ and $L_{\text{max}}=120$ (with
$m_{\text{max}}=30$) (see~\cite{Hol09}), continuation studies of MHD
flows in rotating spherical shells have only reached $N_r=36$ and
$L_{\text{max}}=36$ spectral discretization parameters (see the very
recent study~\cite{FTZH17} where dynamo rotating waves have been
obtained).

\begin{figure}[t!]
\includegraphics[width=0.98\linewidth]{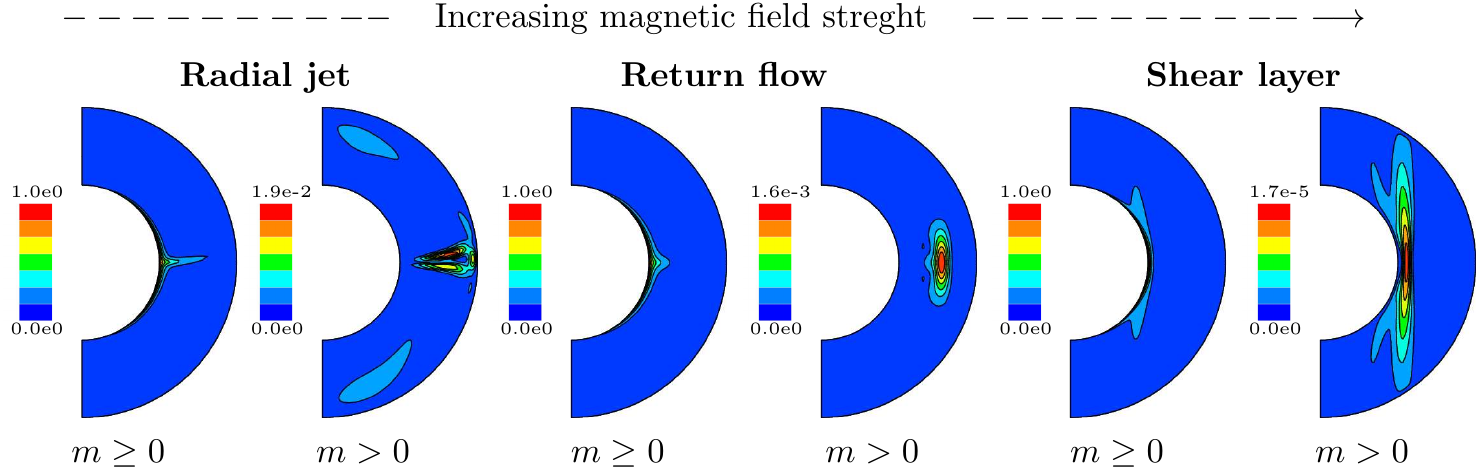}  
\caption{Contour plots of the radial jet ($\Ha=4.32$), return
    flow ($\Ha=29$) and shear layer ($\Ha=79.4$) instabilities. For
    each instability meridional sections (through a relative maximum)
    of $\ve^2/2$ and its nonaxisymmetric component ($m>0$) are
    shown.}
\label{fig:m2_inst} 
\end{figure}

\begin{figure}[b!]
\begin{center}
\includegraphics[width=0.98\linewidth]{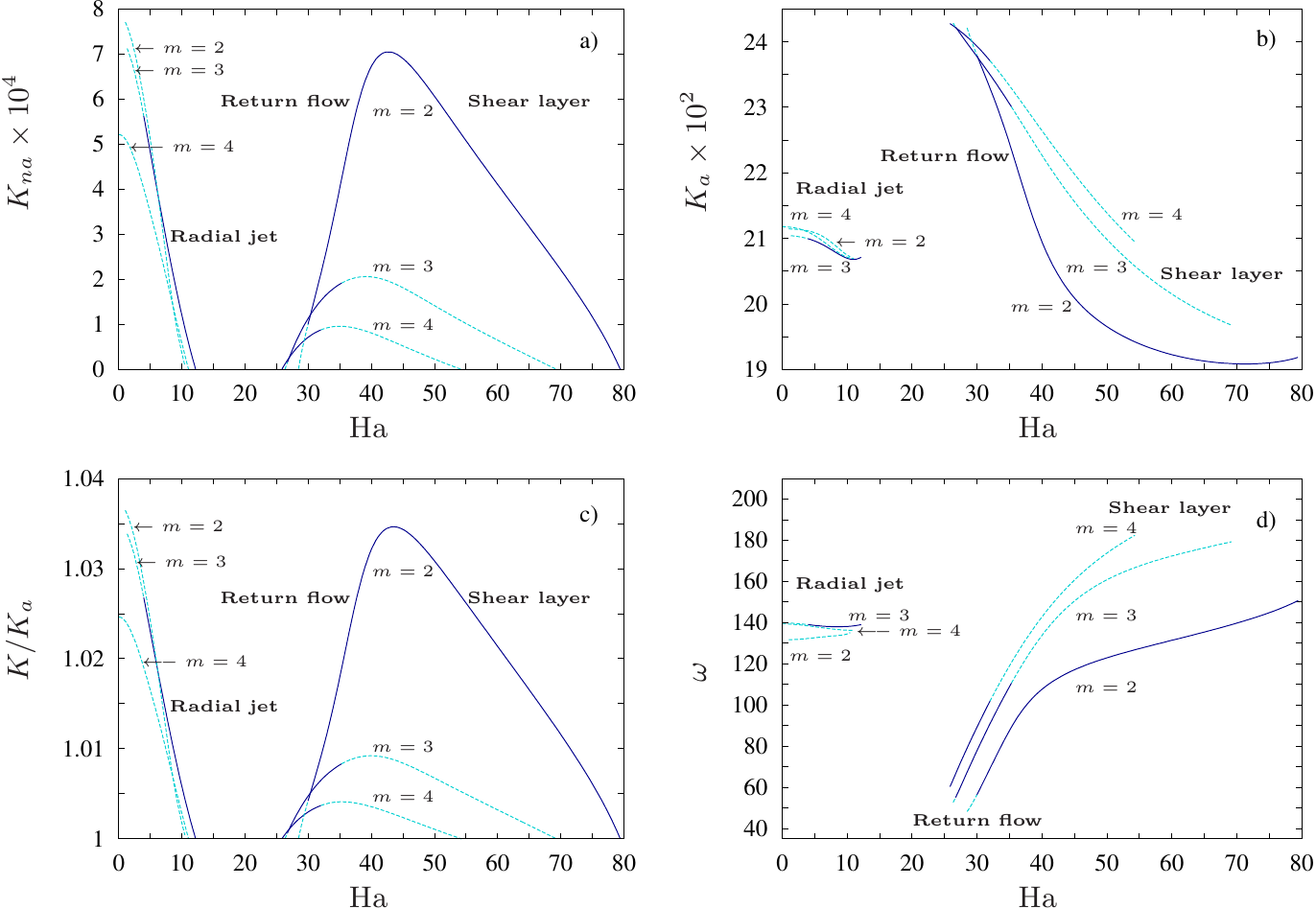}
\end{center}
\caption{Bifurcation diagrams varying $\Ha$ at fixed Reynolds number
  $\Ree=10^3$ and aspect ratio $0.5$ for several time and volume
  averaged properties. (a) nonaxisymmetric kinetic energies $K_{na}$.
  (b) axisymmetric kinetic energies $K_{a}$. (c) Ratios of global to
  axisymmetric energies $K/K_a$. (d) Rotation frequency. Solid/dashed
  lines mean stable/unstable waves. The type of instability is
  labelled in each panel. The radial jet instability (on the left, low
  $\Ha$) is equatorially asymmetric while the return flow or shear
  layer instabilities (on the right, moderate and high $\Ha$) are
  symmetric.}
\label{fig:bif_diagr} 
\end{figure}

\section{Results}
\label{sec:res}

This study constitutes a further step towards the modelling of the
HEDGEHOG experiment~\cite{Kap14,KKSS17}. This study is restricted to
$\eta=0.5$ and $\Ree=10^3$ and $\Ha\in(0,80)$ for which some
experimental data is available~\cite{KKSS17}. For this range of
parameters the linear stability analysis of the basic axisymmetric
state has already been performed in~\cite{TEO11}. At $\Ree=10^3$ and
without magnetic field ($\Ha=0$) the basic state is unstable to
equatorially antisymmetric nonaxisymmetric perturbations developing
the radial jet instability~\cite{HJE06}. This instability, with
azimuthal wave number $m=3$, is maintained by increasing magnetic
strength, but the basic state restabilises again at a critical
Hartmann number $\Ha_c=12.2$ (see Table 3.~\cite{TEO11}). By
increasing $\Ha$ further beyond $\Ha_c=25.8$ another Hopf bifurcation
gives rise to a RW, with $m=4$ and equatorially symmetric, which
corresponds to the the return flow instability~\cite{Hol09}. The
return flow instability is characterised by a meridional circulation
from the equatorial plane in the middle of the shell. At sufficiently
large $\Ha$ the flow instabilities become magnetically confined within
the tangential cylinder and strong shear layers
develop~\cite{HoSk01}. This is the so-called shear-layer
instability. Figure~\ref{fig:m2_inst} exhibits the essential features
of the radial jet, return flow and shear layer nonlinear instabilities
with azimuthal symmetry $m=2$. For the radial jet the nonaxisymmetric
($m>0$) flow is concentrated near the outer boundary and very close to
the equatorial plane. In the case of return flow the $m>0$ flow is
mainly developed in the bulk of the shell, at low latitudes. For the
shear layer the $m>0$ flow elongates on the axial direction, reaching
high latitudes, and it is attached to the inner sphere.

The bifurcation diagrams, i.e branches of RW, for the three types of
instabilities previously described are presented in the
following. Azimuthal wave numbers $m=2,3,4$ are selected because they
are preferred at the onset of the instabilities at aspect ratio
$\eta=0.5$ and $\Ree=10^3$~\cite{TEO11}. They are indeed in
concordance with those that can be measured with ultrasonic Doppler
velocimetry (UDV) probes mounted on the HEDGEHOG
experiment~\cite{KKSS17}. We also describe the flow topology of the
instabilities and report the regions of multistability of the waves
and the critical Hartmann numbers for the secondary bifurcations to
quasiperiodic flows. The symmetry of the latter is identified and some
examples, for both equatorially symmetric and antisymmetric
instabilities, are provided.

\subsection{Bifurcation Diagrams}
\label{sec:res_bif}

Figure~\ref{fig:bif_diagr} contains the branches of RW with azimuthal
wave numbers $m=2,3,4$ versus the Hartmann number $\Ha$, each panel
displaying a different quantity. We recall that a $m=m_d$ nonlinear RW
has an $m_d$-fold azimuthal symmetry (invariance under $2\pi/m_d$
azimuthal rotations) and has nonzero spherical harmonic amplitudes
only for the azimuthal wave numbers $m=km_d$,
$k\in\mathbb{Z}$. Volume-averaged nonaxisymmetric $K_{na}$ and
axisymmetric $K_a$ kinetic energies, and the ratio of total over
axisymmetric volume-averaged kinetic energy $K/K_a$, are represented
in Fig.~\ref{fig:bif_diagr}(a), (b), and (c), respectively. Finally,
the rotation frequency $\omega$ is shown in
Fig.~\ref{fig:bif_diagr}(d). In all the panels, the branches of RW on
the right correspond to the return flow and shear-layer instabilities,
both equatorially symmetric, and those on the left correspond to the
radial jet instability which is equatorially
asymmetric. Unstable/stable RW are denoted with a dashed/solid line.
The general picture of the situation is best displayed in
Fig.~\ref{fig:bif_diagr}(a) (also (c)). Each branch bifurcates from
the axisymmetric basic state (horizontal axis, $K_{na}=0$) at
$\Ha_c^{\text{radial jet}}(m)$, $\Ha_c^{\text{return flow}}(m)$ and
$\Ha_c^{\text{shear layer}}(m)$ with $\Ha_c^{\text{radial
    jet}}(m)<\Ha_c^{\text{return flow}}(m)<\Ha_c^{\text{shear
    layer}}(m)$. The azimuthal wave numbers giving rise to the
critical values are $m=3$, $m=4$ and $m=2$, respectively,
representative of the radial jet, return flow and shear-layer
instabilities. As commented before $\Ha_c^{\text{radial
    jet}}(m=3)=12.2$, $\Ha_c^{\text{return flow}}(m=4)=25.8$
(see~\cite{TEO11}) and we have found $\Ha_c^{\text{shear
    layer}}(m=2)=79.4$. According to bifurcation theory, RW
bifurcating at these critical values are stable, otherwise are
unstable, i.e if they bifurcate from nondominant eigenfunctions.

\begin{figure}[h]
\includegraphics[width=0.98\linewidth]{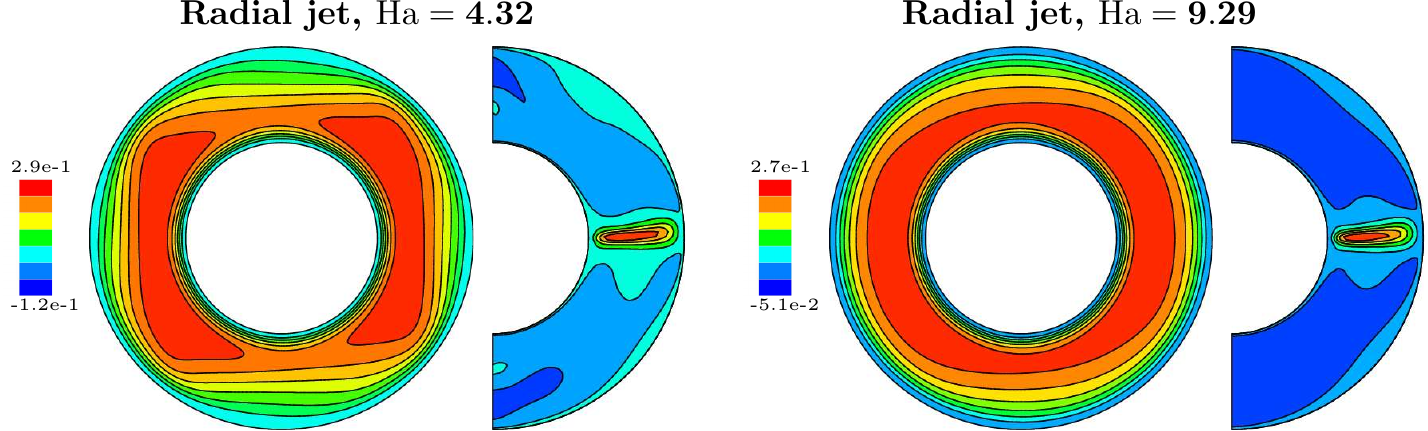}
\caption{Contour plots of equatorially asymmetric rotating
    waves, corresponding to the radial jet instability, on the $m=2$
    branch. The left two plots are the equatorial and meridional
    sections (through a relative maximum) of radial velocity $v_r$ at
    $\Ha=4.32$. Right two plots: Same sections for
    $\Ha=9.29$.}
\label{fig:m2as_co} 
\end{figure}

\begin{figure}[h]
\includegraphics[width=0.98\linewidth]{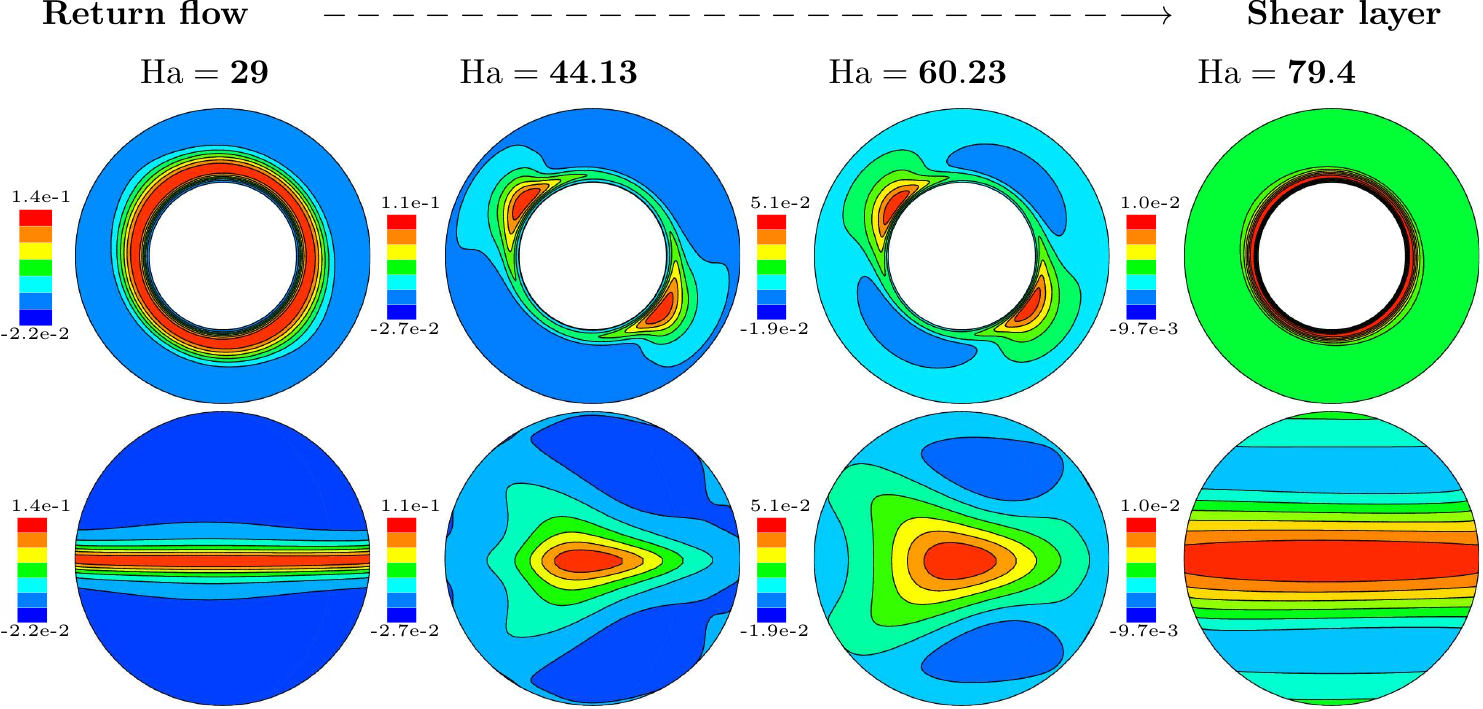}
\caption{Contour plots of equatorially symmetric $m=2$ RW
    showing the continuous transformation from the return flow to the
    shear-layer instability with increasing $\Ha$. Equatorial (first
    row) and spherical (through a relative maximum, second row)
    sections of radial velocity $v_r$.}
\label{fig:m2_co1} 
\end{figure}

At the weakly magnetised regime $\Ha<12.2$, corresponding to the
radial jet instability, only the $m=3$ RW is found stable. All waves
are characterised by a pronounced increase of nonaxisymmetry along the
whole branch but at different rates (the smaller the value of $m$ the
larger the rate), although solutions are still nearly axisymmetric
when $\Ha=0$ (see Fig.~\ref{fig:bif_diagr}(c)). The difference between
kinetic energies or frequencies of the branches is not so large,
especially at $\Ha\in(5,12.2)$ and $K_a$ and $\omega$ vary smoothly
with $\Ha$. The characteristic flow topology of these radial jet
instabilities, exhaustively described in~\cite{HJE06}, can be seen in
Fig.~\ref{fig:m2as_co}. Figure~\ref{fig:m2as_co} displays a equatorial
and meridional section of the radial velocity $v_r$ (1st and 2nd
plots, from left to right) for a RW with $m=2$ azimuthal wave number
and $\Ha=4.32$. The same sections are on the right group of two plots
but for the $m=2$ RW at $\Ha= 9.29$. The flow results in an equatorial
radial jet which emerges from the inner boundary. The flow's
meridional circulation is enhanced when the radial jet reaches the
outer boundary. Because of the drifting nature of the nonaxisymmetric
flow the time evolution of the pattern is seen as an oscillation of
the jet around the equatorial plane. The larger the nonaxisymmetric
component (i.e the smaller $\Ha$), the higher latitudes the flow can
affect (compare meridional sections of $v_r$ in
Fig.~\ref{fig:m2as_co}). The study of the location of convection is
important from an experimental point of view since it helps to decide
on the optimal position of measurement probes.

\begin{figure}[h!]
\includegraphics[width=0.98\linewidth]{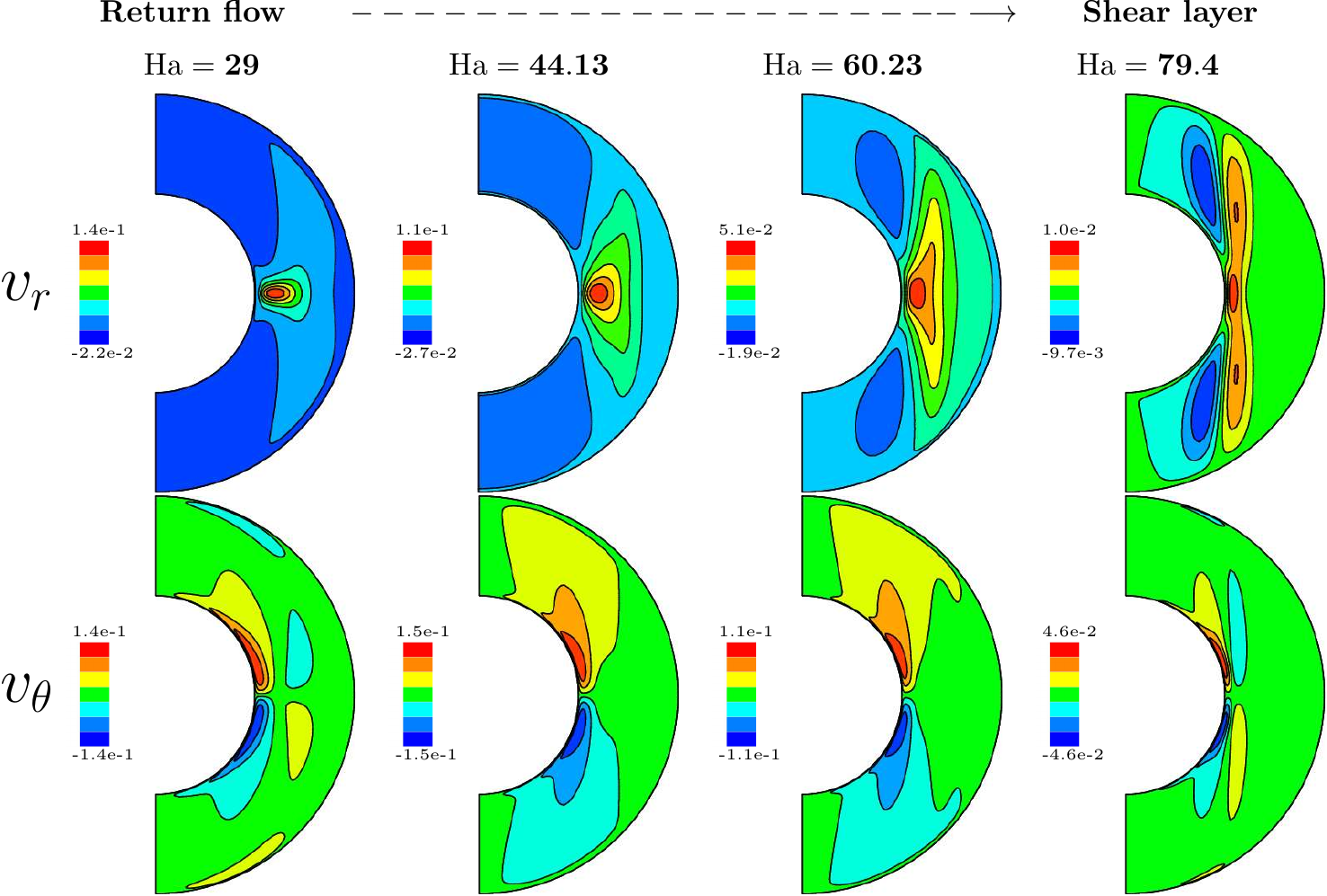}  
\caption{Contour plots of equatorially symmetric $m=2$ RW
    showing the continuous transformation from the return flow to the
    shear-layer instability with increasing $\Ha$. Meridional sections
    (through a relative maximum) of the radial $v_r$ (first row) and
    colatitudinal $v_{\theta}$ (second row) velocities are shown.}
\label{fig:m2_co2} 
\end{figure}

An equatorially symmetric (return flow instability) $m=4$ RW emerges
at $\Ha=25.8$ (see Fig.~\ref{fig:bif_diagr}(a)). This is quite close
to $\Ha=26.3$, corresponding to the bifurcation of the $m=3$ RW.
Then, a double-Hopf bifurcation similar to that described
in~\cite{SGN13} for a thermal convection in spherical geometry problem
could be found by moving a second parameter ($\Ree$ or $\eta$). At
$\Ha=28.5$ the remaining $m=2$ RW branch starts. Because the unstable
$m=2$ and $m=3$ RW become stable very soon, especially in the case of
$m=3$ RW, several regions of multistability arise. They correspond to
the return flow instability and will be accurately computed in
\S\ \ref{sec:res}\ref{sec:res_stab}. From the beginning of the return
flow instability $K_{na}$ and $\omega$ increase sharply whilst $K_{a}$
decreases for all the branches. In contrast to what occured in the
radial jet instability, the differences between $K_{na}$ or $K_{a}$ of
the $m=2,3,4$ branches start to rise significantly for $\Ha>30$.  The
same applies to the difference between $\omega$ values for
$\Ha>40$. Detecting these differences is of experimental relevance
because it helps to select the proper azimuthal wave number (that with
a maximum signal) and thus guide in the measurement design, by
positioning accordingly the UDV.

The return flow instability of the basic state turns gradually, by
increasing $\Ha$, into the shear-layer instability~\cite{Hol09}. This
also occurs when nonlinear effects are included giving rise to
equatorially symmetric RW.  For each $m$, the branch starts and ends
at the basic state. Hence there is some $\Ha^*$ which maximises
$K_{na}$ as shown in Fig.~\ref{fig:bif_diagr}(a). We have found that
close to this $\Ha^*$ the instability can no longer be strictly
classified as return flow or shear-layer because it contains features
of both. Then, it seems reasonable to consider this $\Ha^*$ as the
critical Hartmann number defining the boundary between return flow and
shear-layer instabilities. Notice in Fig.~\ref{fig:bif_diagr}(a) that
$\Ha^*$ can be defined for each branch, i.e,
$\Ha^*\equiv\Ha^*(m)$. According to our results
$\Ha^*(m_1)>\Ha^*(m_2)$ and
$K^{\text{max}}_{na}(m_1)>K^{\text{max}}_{na}(m_2)$ provided that
$m_1<m_2$ (the maximum value of the nonaxisymmetric kinetic energy in
each branch $m$ is $K^{\text{max}}_{na}(m)$). Then, the smaller the
value of $m$ the larger the value of $K^{\text{max}}_{na}(m)$ which
again is interesting for experimental purposes: Nonaxisymmetric
signals ($m>0$) may be maximised by choosing properly $\Ha$ close to
$\Ha^*(m)$. In addition, mounting sensors to measure low azimuthal
wave number properties may also give rise to a larger
  signal. The existence of this critical $\Ha^*$ separating two
different flow behaviours can also be inferred from the frequency
dependence seen in the branches of
Fig.~\ref{fig:bif_diagr}(d). Approximately at $\Ha^*$ there is a
change of slope (best seen for $m=2$), indicating that frequencies
change smoothly when shear-layer instabilities are selected.

Figure~\ref{fig:bif_diagr}(a) shows that the smaller the $m$ the
larger is the corresponding equatorially symmetric branch. This is in
correspondence with the shape of the marginal stability curves
corresponding to the basic state shown in Fig. 4
of~\cite{TEO11}. Looking at this figure and comparing with
Fig.~\ref{fig:bif_diagr}(a) we could imagine the $m=5$ and $m=6$
branches of RW to lie just below the $m=4$ branch having both very
small $K_{na}$. On the other hand, because all branches terminate at
the basic state, Fig.~\ref{fig:bif_diagr}(b) provides some information
about the basic axisymmetric flow. This figure shows how the amplitude
of the axisymmetric flow is increased from radial jet to return flow
instability but is strongly decreased when shear-layer instabilities
are selected at the higher $\Ha$.

The features of the equatorially symmetric RW described above can be
better visualised with the help of contour plots. The first row of
Fig.~\ref{fig:m2_co1} displays the contour plots of the radial
velocity $v_r$ on the equatorial plane for a sequence of $m=2$ RW with
increasing Hartmann number (from left to right $\Ha=29$, $\Ha=44.13$,
$\Ha=60.23$, and $\Ha=79.4$). The left and right equatorial sections
correspond to the return flow and shear-layer instabilities,
respectively. The equatorial sections in between show the continuous
transformation between both instabilities and the corresponding
increase and decrease of the amplitude of nonaxisymmetric
flow. Spherical sections (at constant $r$) of $v_r$ are given in the
2nd row. Their radial positions correspond to those of the maximum,
$r=r_i+\alpha d$ with $\alpha=0.2,0.19,0.15,0.12$ (from left to
right). For the return flow instability the position of maximum $v_r$
becomes fairly constant (three left spherical sections, $\Ha< 45$) and
increasing $\Ha$ results in a wider radial jet which spreads to higher
latitudes. Beyond $\Ha^*\approx 45$ the RW should be classified as
shear-layer instability. The radial jet is progressively stretched to
the inner boundary reaching higher latitudes.  A view of the
meridional sections of radial $v_r$ and colatitudinal $v_{\theta}$
velocity of Fig.~\ref{fig:m2_co2} (1st and 2nd rows, respectively),
completes the picture of the transformation of return flow into the
shear-layer instability. Again the sections cut a maximum of each
component of the velocity. The stretching of the return flow
instability is clearly seen on the sections of $v_r$ (first row). The
latter also depicts the development of strong shear layers, and the
progressive alignment of the flow with the rotation axis, when the
corresponding instability is selected (right sections, $\Ha> 45$). The
sections of $v_{\theta}$ (2nd row) confirm this tendency. For the
shear-layer instability the strong magnetic field inhibits convection
outside the tangent cylinder (parallel to the rotation axis and
tangent to the inner sphere). This is because the fluid in this region
is only coupled with the outer boundary which is at rest whereas
within the tangent cylinder the fluid is affected also by the inner
moving boundary.

\subsection{Stability Analysis: Transition to Quasiperiodic Flows}
\label{sec:res_stab}

According to bifurcation (Floquet) theory a periodic orbit is stable
for a given $\Ha$ as long as all its Floquet multipliers
$\lambda_k\in\mathbb{C}$ lie within the unit circle $|\lambda_k|<1$,
$k=1,...,n$. When, by varying $\Ha$, one of these multipliers crosses
the unit circle the periodic orbit becomes unstable. That multiplier
gives information about the type of bifurcation and the type of
solutions of the branch emerging at the critical point. For instance
if $\lambda=\lambda_r+i\lambda_i$ is real ($\lambda_i=0$) then a
pitchfork or period doubling bifurcation, giving rise to periodic
orbits of the same or twice the period of the RW, respectively, occurs
depending on whether $\lambda_r=1$ or $\lambda_r=-1$, respectively. If
by contrast $\lambda$ is a complex eigenvalue with $\lambda_i\ne0$ the
bifurcation is of Hopf type and the new solution branch is an
invariant torus, i.e, a quasiperiodic orbit with two frequencies. The
quasiperiodic scenario is found in all branches of RW, either
equatorially asymmetric or symmetric.

In the absence of a magnetic field, $\Ha=0$, the critical Reynolds
number is around $\Ree_c\approx 489$ (see~\cite{HJE06}). Because we
have chosen a significantly supercritical Reynolds of $\Ree=10^3$,
rotating waves with $\Ha\sim 0$ are expected to be unstable. This is
what occurs in the $m=2,3,4$ branches of radial jet instability which
have several multipliers outside the unit circle. Because of the
stabilising effect of the magnetic field the unstable multipliers
eventually cross the unit circle giving rise to Hopf bifurcations. In
the case of $m=2$ or $m=4$ RW there is always one unstable multiplier
whereas the $m=3$ RW stabilises at $\Ha_c=3.95$ until the critical
Hartmann for the basic state is reached ($\Ha=12.2$). At the critical
parameter $\Ha_c=3.95$ the the azimuthal symmetry of the eigenfunction
is $m=1$ (i.e invariant by $2\pi$ azimuthal rotations) (see
Table~\ref{table:bif_asym}) and then a branch of modulated rotating
waves MRW with azimuthal symmetry $m=1$ is born. Whether MRW are
stable or not depends on whether the bifurcation is subcritical or
not. It also depends on whether the parent RW state from which the MRW
bifurcates is stable or not. MRWs bifurcating from unstable RWs will
also be unstable. We have found that the bifurcation is subcritical
and then MRW are stable with $\Ha\lesssim \Ha_c$. An example of MRW
obtained with DNS will be shown later on. In addition, for $\Ha
\lesssim 3.33$ the $m=3$ RW's have a 2nd unstable eigenfunction with
azimuthal symmetry $m=3$ (invariant to $2\pi/3$ azimuthal
rotations). Then, at $\Ha_c=3.33$ an unstable branch of MRW with
azimuthal symmetry $m=3$ is born. Detecting the existence of this
branches is important for a deeper understanding of temporal chaotic
flows in the range $\Ha\in[0,5]$.

\begin{table}[ht] 
  \begin{center}
    \begin{tabular}{lcccccccc}
\vspace{0.1cm}            
& $m$  & Tran.         & $\Ha_c$ & $\omega$ & $\text{Arg}(\lambda)$ & Azim. symm. $m$ \\
\hline\\
& $2$  &  $(+,-)$ to $(+,+)$  & $9.43$ & $133.82$ & $0.10$                 & $2$\\
& $3$  &  $(+,-)$ to $(-,-)$  & $3.95$ & $139.07$ & $0.77$                 & $1$\\
& $4$  &  $(+,-)$ to $(+,+)$  & $9.97$ & $136.08$ & $1.49$                 & $1$\\
& $4$  &  $(+,+)$ to $(+,-)$  & $9.98$ & $136.08$ & $3.13$                 & $2$\\
\hline
  \end{tabular}
  \caption{Critical parameters of the asymmetric (radial jet) RWs at
    the bifurcations where they change the stability
    ($|\lambda|=1$). These parameters are obtained by inverse
    interpolation with a polynomial of degree $5$. Transitions are
    denoted by Tran. The sign of the Floquet exponent of the two
    dominant eigenfunctions before and after the transition is
    shown. The azimuthal symmetry ($2\pi/m$ azimuthal periodicity) of
    the eigenfunction with a change of sign is also stated.}
  \label{table:bif_asym}
  \end{center}
\end{table}

\begin{figure}[t!]
\includegraphics[width=0.98\linewidth]{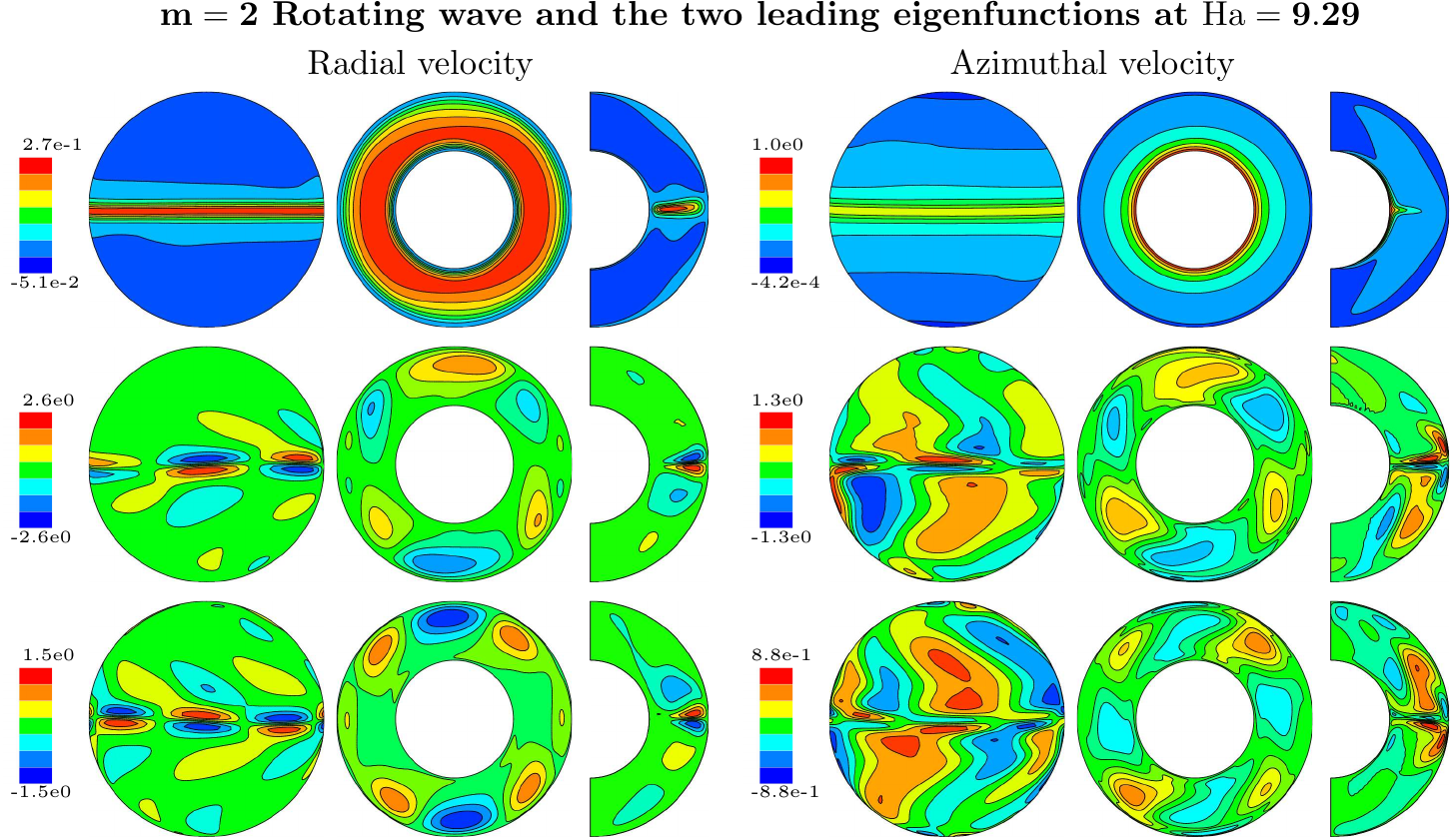}    
\caption{Contour plots of an $m=2$ unstable equatorially asymmetric
  rotating wave (first row) and its two dominant eigenfunctions
  (second and third rows) at $\Ha=9.29$ corresponding to the radial
  jet instability. First row: the left three plots are the spherical,
  equatorial and meridional sections of the radial velocity
  $v_r$. Spherical and meridional sections are taken where $v_r$ has a
  relative maximum. Right three plots: Same sections for
  $v_{\varphi}$. The spherical section is at $r=r_i+0.1d$ and the
  meridional section cuts a relative maximum. Second/third row: As
  first row but for the first/second dominant eigenfunctions,
  respectively. In this case the spherical and meridional sections are
  taken at a relative maximum.}
\label{fig:m2as} 
\end{figure}

\begin{figure}[t!]
\includegraphics[width=0.98\linewidth]{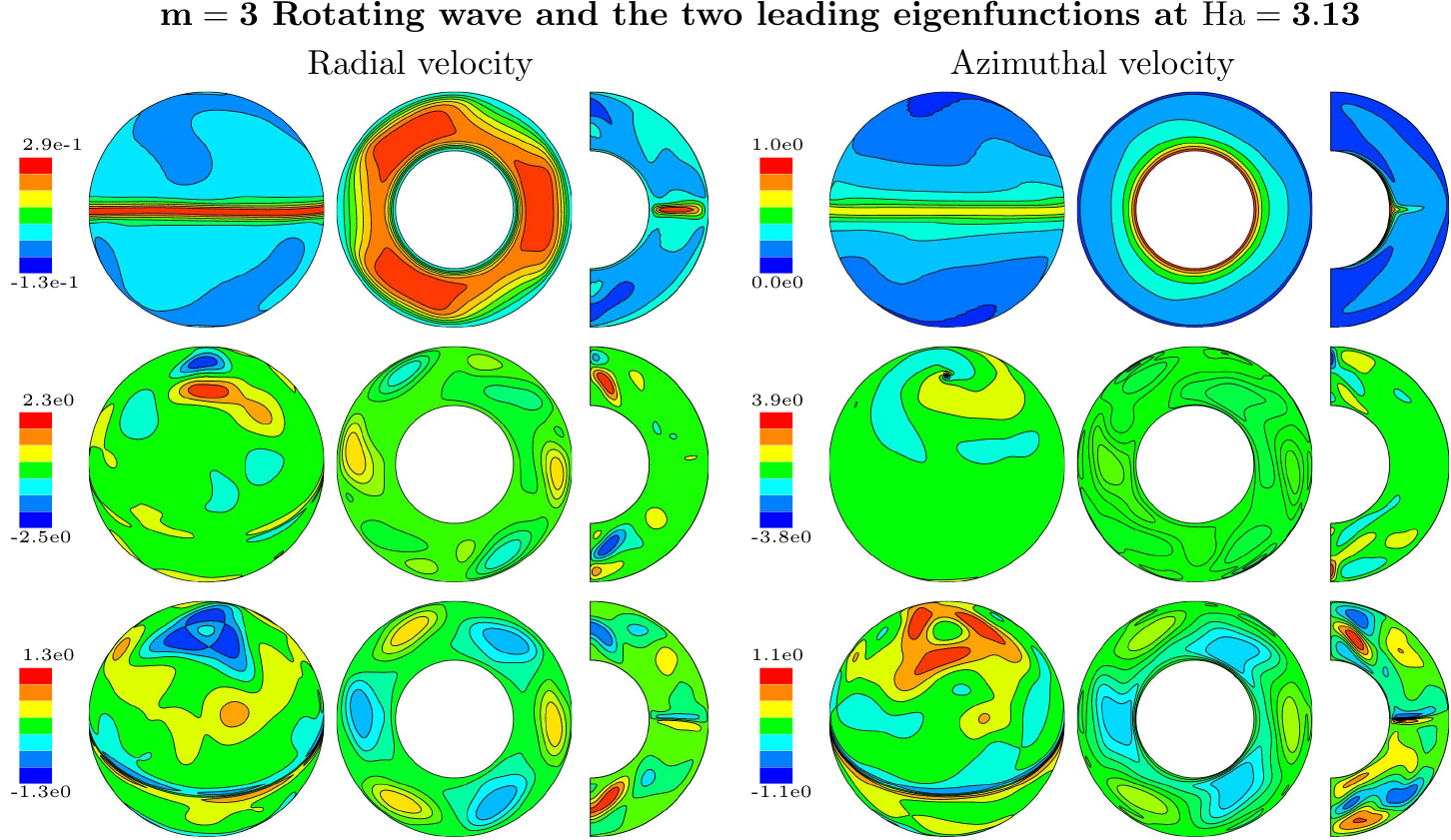}      
\caption{As Fig.~\ref{fig:m2as} but for an $m=3$ RW at
  $\Ha=3.13$.}
\label{fig:m3as} 
\end{figure}

For all $\Ha<12.2$ explored the $m=2$ RW's are unstable with dominant
eigenfunction with azimuthal symmetry $m=1$. At $\Ha \gtrsim 9.43$ the
$m=2$ RW's have a 2nd unstable eigenfunction with azimuthal symmetry
$m=2$ (invariant to $2\pi/2$ azimuthal rotations) and thus a branch of
unstable MRW with azimuthal symmetry $m=2$ could be found near
this range. Since these MRW retain the $m=2$ symmetry they can be
found with $m=2$ azimuthally constrained DNS. A similar behaviour is
found for the $m=4$ RW. For $\Ha \lesssim 9.98$ they are unstable with
an eigenfunction with azimuthal symmetry $m=2$. In addition,
for $\Ha \gtrsim 9.97$ the $m=4$ RW's are unstable with an
eigenfunction with azimuthal symmetry $m=1$. Then, for
$\Ha\in(9.97,9.98)$ the $m=4$ RW have two unstable eigenfunctions,
while for $\Ha$ barely outside of this interval RW's have only one
unstable eigenfunction. We estimate $\Ha\approx9.98$ to be the locus
of a double-Hopf bifurcation.

\begin{figure}[t!]
\includegraphics[width=0.98\linewidth]{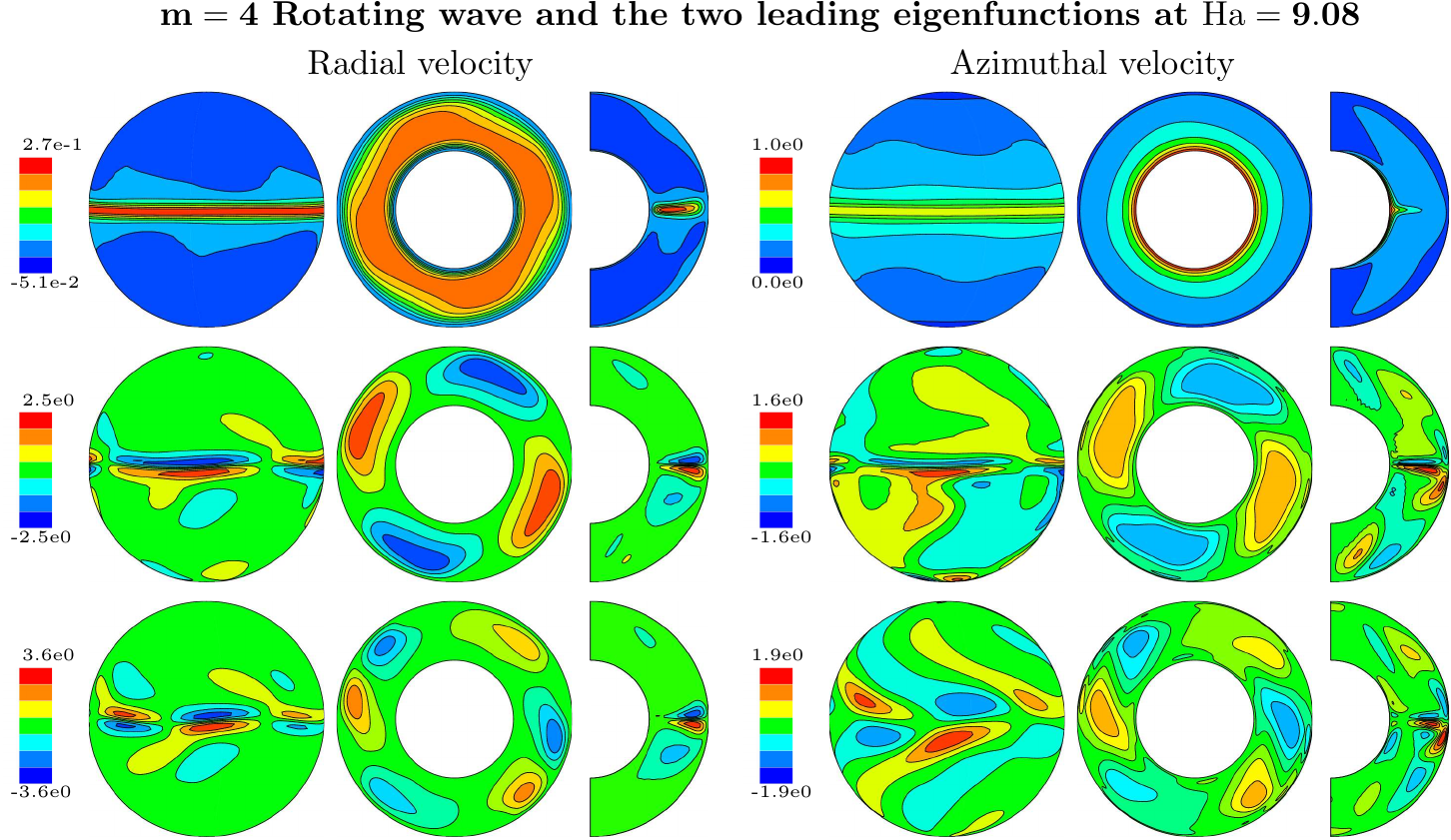}      
\caption{As Fig.~\ref{fig:m2as} but for an $m=4$ RW at
  $\Ha=9.08$.}
\label{fig:m4as} 
\end{figure}

Figures~\ref{fig:m2as}, \ref{fig:m3as}, and~\ref{fig:m4as} display the
patterns (radial and azimuthal velocity) of RW and their dominant
eigenfunctions lying on the $m=2$, $m=3$ and $m=4$ branches,
respectively. The values of $\Ha$ are selected close to bifurcation
points. While the patterns of the different $m$ RW remain quite
similar, those of the eigenfunctions appear to be noticeably
different. For small magnetic forcings $\Ha<5$ the instability giving
rise to MRW is mainly concentrated within the tangent cylinder, i.e,
in the polar regions, see Fig.~\ref{fig:m3as} corresponding to a $m=3$
RW at $\Ha=3.13$. Because these $m=3$ RW are equatorially
asymmetric and their dominant eigenfunction has azimuthal
  symmetry $m=1$, the bifurcated MRW will have all the spatial
symmetries broken. In the context of convection in spherical shells
these solutions are quite rare and have been never reported before. In
contrast, for larger values of $\Ha$ the instabilities lie closer
to the outer surface in the equatorial plane. Nevertheless, the
instability also affects higher latitudes as it reflects the spherical
sections of $v_{\varphi}$ shown in Figs.~\ref{fig:m2as}
and~\ref{fig:m4as}, with the $m=2$ instabilities being the most
altered in this region.

\begin{table}[b!] 
  \begin{center}
    \begin{tabular}{lcccccccc}
\vspace{0.1cm}            
& $m$ & Tran. & $\Ha_c$ & $\omega$ & $\text{Arg}(\lambda)$ & Azim. symm. $m$ \\
\hline\\
& $2$ &  $(+,-)$ to $(-,-)$ & $30.04$ & $56.83$  & $0.31$                & $1$  \\
& $3$ &  $(-,-)$ to $(-,-)$ & $26.68$ & $54.98$  & $2.52$                & $1$ \\
& $3$ &  $(-,-)$ to $(+,-)$ & $35.24$ & $110.36$ & $2.83$                & $1$ \\
& $4$ &  $(-,-)$ to $(+,-)$ & $31.95$ & $101.62$ & $2.077$               & $1$  \\
\hline
  \end{tabular}
  \caption{Critical parameters of the symmetric RWs at the
    bifurcations where they change the stability
    ($|\lambda|=1$). These parameters are obtained by inverse
    interpolation with a polynomial of degree $5$. Transitions are
    denoted by Tran. The sign of the Floquet exponent of the two
    dominant eigenfunctions before and after the transition is
    shown. The azimuthal symmetry ($2\pi/m$ azimuthal periodicity) of
    the eigenfunction with a change of sign is also stated.}
\label{table:bif_sym}
\end{center}
\end{table}

\begin{figure}[t!]
\begin{center}
\includegraphics[width=0.6\linewidth]{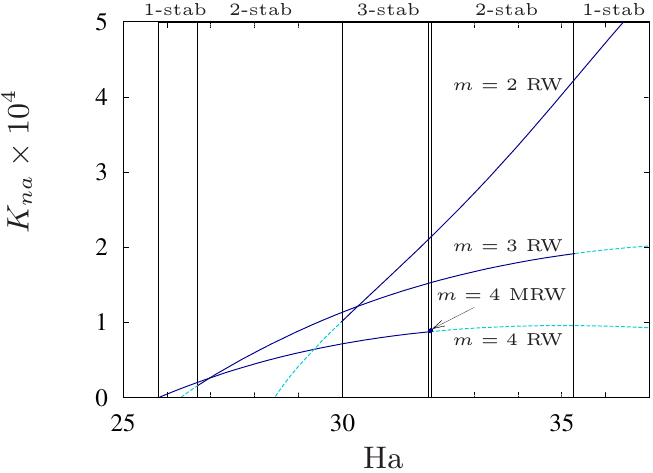}
\end{center}
\caption{Bifurcation diagram varying $\Ha$ at fixed Reynolds
    number $\Ree=10^3$ and aspect ratio $0.5$ showing the time and
    volume averaged nonaxisymmetric kinetic energies $K_{na}$ for the
    return flow instabilities. Solid/dashed lines mean stable/unstable
    waves. The different regions of multistability are highlighted
    with bands: one stable solution (cross-hatched), bi-stability
    (small cross-hatched) and tri-stability (solid). The narrow region
    corresponds to the stability region of MRW.}
\label{fig:bif_diagr_sym} 
\end{figure}

The stability analysis of the return flow and shear-layer type
equatorially symmetric RW, summarised in table~\ref{table:bif_sym},
reveals several regions of multistability. These regions are
displayed in Fig.~\ref{fig:bif_diagr_sym}. By means of DNS
explorations we have found that the Hopf bifurcation where $m=4$ RW
lose the stability is supercritical. Then, there exists a small
interval to the right of $31.95$ over which an MRW is stable giving
rise to a region of tri-stability (the thinner solid yellow band of
Fig.~\ref{fig:bif_diagr_sym}). The eigenfunctions at the bifurcation
point have $m=1$ azimuthal symmetry and so do the bifurcated MRW. This
also occurs at the bifurcations on the $m=2$ and $m=3$ branches.

\begin{figure}[h!]
\includegraphics[width=0.98\linewidth]{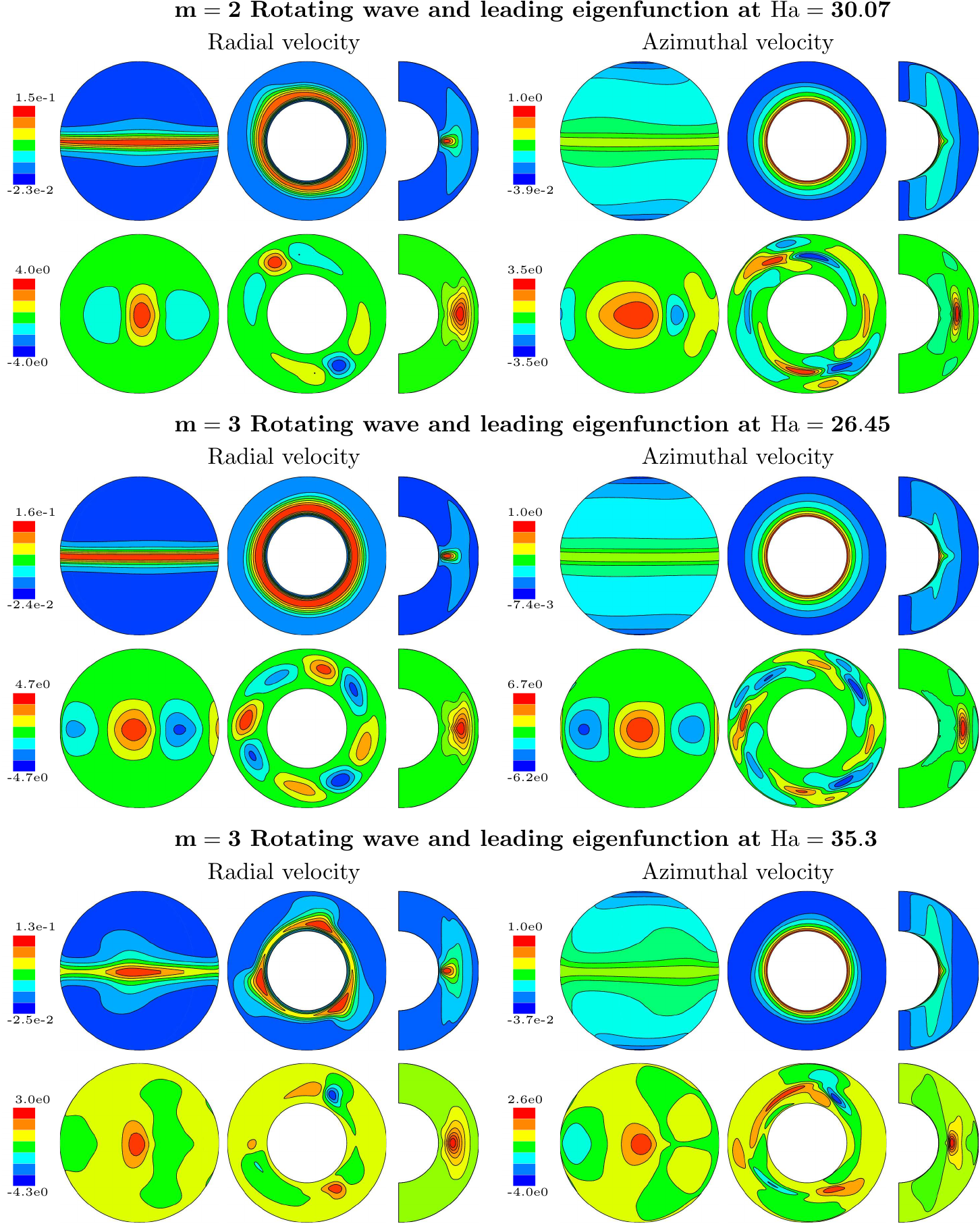}        
\caption{Contour plots of $m=2,3,4$ equatorially symmetric
    rotating waves and their dominant eigenfunctions close to
    bifurcation points corresponding to the return flow
    instability. For each $m$ the left three plots are the spherical,
    equatorial and meridional sections of the radial velocity $v_r$
    and the right three plots are the same sections for
    $v_{\varphi}$.}
\label{fig:all_m_sym} 
\end{figure}

The RW and their dominant eigenfunction patterns close to the
bifurcation points of table~\ref{table:bif_sym}, from top to bottom,
are displayed in Fig.~\ref{fig:all_m_sym}. They are all representative
of return flow instability because the bifurcations occur at
relatively moderate magnetic forcings $\Ha<45$. In all the cases the
eigenfunctions are equatorially symmetric and the instability is
concentrated in the bulk of the shell, specifically located on the
edge of the radial jet. Notice that although the azimuthal symmetry is
$m=1$, in the case of $m=2$ and $m=4$ branches the eigenfunctions
satisfy some additional symmetry, namely, they are invariant under
azimuthal rotations of $\pi$ degrees and change of sign. This does not
occur on the $m=3$ branch where the azimuthal symmetry is strictly
$m=1$, i.e only invariant under $2\pi$ azimuthal rotations.

\begin{figure}[t!]
\begin{center}
\includegraphics[width=0.95\linewidth]{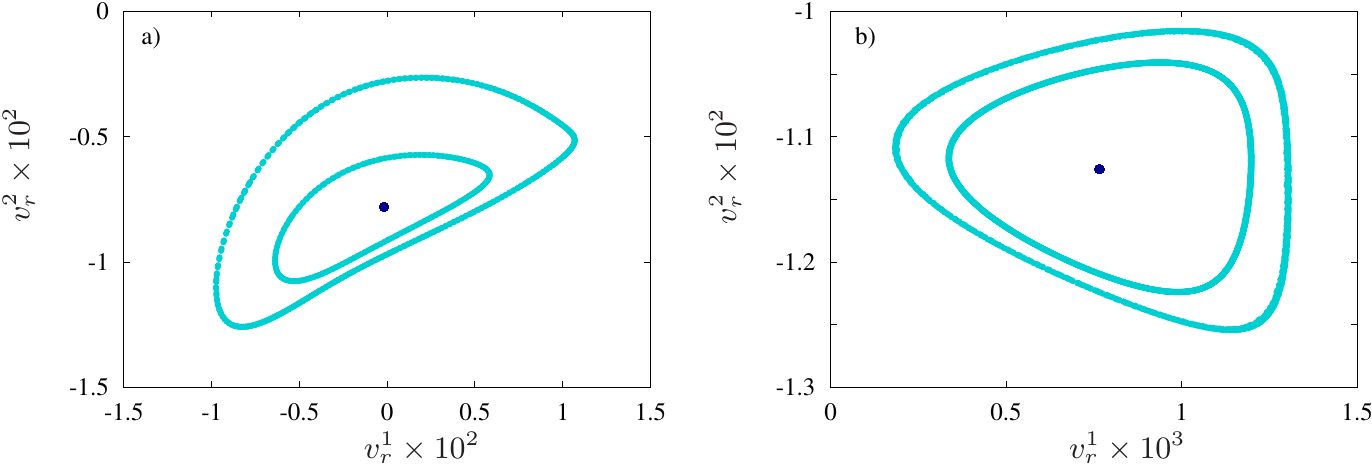}
\end{center}
\caption{Poincar\'e sections of the radial jet and return flow
  instabilities (a) The Poincar\'e section is defined by
  $v_r(r_s,\theta_s,\varphi_s)=0$, with
  $(r_s,\theta_s,\varphi_s)=(1.5,3\pi/4,0)$. Invariant tori at
  $\Ha=3.80$ (inner closed curve) and at $\Ha=3.54$ (outer
  closed curve) bifurcated from the $m=3$ asymmetric branch. The inner
  point corresponds to a stable $m=3$ periodic orbit at
  $\Ha=4.19$. (b) The Poincar\'e section is defined by
  $v_r(r_s,\theta_s,\varphi_s)=0$, with
  $(r_s,\theta_s,\varphi_s)=(1.854,5\pi/8,0)$. Invariant tori at
  $\Ha=32$ (inner closed curve) and at $\Ha=32.02$ bifurcated from the
  $m=4$ symmetric branch. The inner dot corresponds to a stable $m=3$
  periodic orbit at $\Ha=31.9$. The axis are defined by
  $v_r^i=v_r(r_i,\theta_i,\varphi_i),\quad i=1,2$ with
  $(r_1,\theta_1,\varphi_1)=(1.5,5\pi/8,0)$ and
  $(r_2,\theta_2,\varphi_2)=(1.146,5\pi/8,0)$. }
\label{fig:psec} 
\end{figure}

As previously mentioned we have found, by means of time-stepping the
model equations close to their bifurcation points, two branches of
stable MRW. Specifically, one of those branches bifurcates
subcritically from the equatorially asymmetric $m=3$ RW at $\Ha=3.95$,
characteristic of the radial jet instability, and the other bifurcates
supercritically from the equatorially symmetric $m=4$ RW at
$\Ha=31.95$, characteristic of the return flow instability. Our
explorations with DNS suggest that the interval of stability of these
MRW is quite small and because of this very long initial transients
are required to saturate the instability, even with initial conditions
that are very close to the final state. This makes finding MRW by
means of DNS quite challenging and this is why very few MRW, have been
reported previously in the literature for this problem
(see~\cite{Hol09} for instance). The computation of MRW branches is
important because they give rise to three-frequency
solutions~\cite{GNS16}, which are the last stage in the route to
turbulence~\cite{RuTa71}.

Quasiperiodic MRW are easily identified because their
azimuthally-averaged properties are periodic, i.e, one frequency
corresponds to the azimuthal drift of the wave and the other to the
modulation. However, in some situations the modulation frequency could
undergo period doubling bifurcations (see~\cite{GSDN15} for instance)
making it difficult to identify the quasiperiodic character of the
solution, even with the help of a frequency spectrum. For a better
classification, Poincar\'e sections provide additional information. By
means of Poincar\'e sections, periodic orbits are represented by a
single point and quasiperiodic solutions as a closed curve, with loops
if a period doubling bifurcation has occured~\cite{GSDN15}. The
Poincar\'e sections of the MRW bifurcated from the $m=3$ RW branch at
$\Ha=3.95$ and the MRW bifurcated from the $m=4$ RW branch at
$\Ha=31.95$, characteristic of the radial jet and return flow
instability, respectively, are plotted in Fig.~\ref{fig:psec}(a) and
(b), respectively. The sections are defined by means of the radial
velocity $v_r(r_s,\theta_s,\varphi_s)=0$, where
$(r_s,\theta_s,\varphi_s)=(1.5,3\pi/4,0)$ is a point in the middle of
the shell at mid latitudes. On both panels the horizontal axis
represents $v_r(1.5,5\pi/8,0)$ and the vertical axis
$v_r(1.146,5\pi/8,0)$. On Fig.~\ref{fig:psec}(a) the point represents
a stable RW at $\Ha=4.19$. Because the bifurcation is subcritical, by
decreasing $\Ha$, a branch of MRW with azimuthal symmetry $m=1$ is
found. The two closed curves of Fig.~\ref{fig:psec}(a) are
representative of solutions belonging to this branch. The inner curve
corresponds to $\Ha=3.80$ and the outer to $\Ha=3.54$. Analogously in
Fig.~\ref{fig:psec}(b) the two closed curves are MRW, but in this case
the region of stability is smaller. The point corresponds to
$\Ha=31.9$, the inner curve to $\Ha=32$ and the outer to $\Ha=32.02$.

\section{Summary}
\label{sec:sum}

The present analysis constitutes a further step towards a better
understanding of the weakly nonlinear regime of the axially magnetised
spherical Couette problem by extending previous
studies~\cite{Hol09,TEO11,Kap14} based on very few nonlinear solutions
and on sketches of bifurcation diagrams. Our study represents
  the first application of Newton-Krylov continuation and stability
  analysis of the solutions to the MSC problem. For such high
dimensional systems, modelling three-dimensional flows, there exist
very few studies based on continuation methods, even in the more
general context of fluid dynamics~\cite{DWCDDEGHLSPSSTT14,SaNe16}.

Thanks to the use of continuation techniques a large number of
nonlinear rotating waves have been obtained and the stability of
around one third of the solutions have been established in a parameter
regime of experimental interest, i.e Reynolds number $\Ree=10^3$,
aspect ratio $\eta=0.5$ and Hartmann numbers $\Ha\in(0,80)$
corresponding to the working conditions of the HEDGEHOG experimental
device at the Helmholtz-Zentrum Dresden-Rossendorf~\cite{KKSS17}. In
this regime radial jet, return flow and shear-layer instabilities have
been previously described~\cite{TEO11,Kap14}.

For a fixed Reynolds number we have accurately determined several
regions of multistability and several transitions to quasiperiodic
flows (modulated waves) of the first nonaxisymmetric instabilities
with azimuthal wave numbers $m=2,3,4$ that occur when the Hartmann
number is varied. The symmetries of the bifurcated MRWs are identified
and some examples are provided. The RW and their eigenfunctions at the
bifurcation points provide initial conditions for the continuation of
MRW that could be performed as in~\cite{GNS16}.

The patterns, rotation frequencies (i.e time scales) and some physical
properties (nonaxisymmetric kinetic energy, for instance) of the
radial jet, return flow and shear-layer instabilities have been
exhaustively described in the range of $\Ha\in(0,80)$. For each
azimuthal wave number the return flow instability changes continuously
with increasing $\Ha$ to become the shear-layer
instability. Both instabilities belong to the same branch. At
the boundary between both instabilities, defined by a critical $\Ha$,
the nonaxisymmetric part of the flow is maximum. Nonaxisymmetry
increases with decreasing wave number $m$. The critical $\Ha$ where
nonaxisymmetric flow is maximum increases with decreasing $m$, as
well.

The determination of stability regions is crucial for comparison with
experiments. They allow, for instance, to determine the azimuthal wave
number of the most physically realisable solutions and thus help to
design appropriate measurement set-ups. The azimuthal wave numbers of
the branches chosen for this study matches with those that can be
measured in the experiments~\cite{KKSS17}. The analysis of physical
properties, such as the volume-averaged kinetic energies, serves as a
prior estimate of their experimental values and thus as a guide for
tuning measurement techniques. Once the solutions are obtained with
continuation methods they can later be easily processed to obtain
other measurable properties not shown in this study, for instance
local velocities inside the shell or the torque acting on the outer
sphere, for a further comparison with experiments.\vskip6pt






\section{Acknowledgements}

F. Garcia was supported by a postdoctoral fellowship of the Alexander
von Humboldt Foundation hosted by Helmholtz-Zentrum
Dresden-Rossendorf. The authors thank the anonymous reviewers for
constructive comments.



\end{document}